\documentclass[twocolumn,showpacs,pre,10pt]{revtex4-1}
\usepackage[english]{babel}
\usepackage{amsmath,amssymb}
\usepackage{times}
\usepackage{epsfig}
\usepackage{ulem}
\usepackage[colorlinks,linkcolor = blue,citecolor = blue]{hyperref}
\usepackage[usenames,dvipsnames]{xcolor}

\newcommand{\MBQ}{many-body quantum }
\newcommand{\ME}{matrix elements }
\newcommand{\I}{integrable }
\newcommand{\NI}{non-integrable }
\begin{document}
\title{Eigenstate thermalization within isolated spin-chain systems}
\author{R. Steinigeweg$^{1}$}
\author{J. Herbrych$^{1}$}
\author{P. Prelov\v{s}ek$^{1,2}$}
\affiliation{$^1$J. Stefan Institute, SI-1000 Ljubljana, Slovenia}
\affiliation{$^2$Faculty of Mathematics and Physics, University of Ljubljana, SI-1000 Ljubljana, Slovenia}
\date{\today}
%05.60.Gg Quantum transport
%71.27.+a Strongly correlated electron systems; heavy fermions
%75.10.Pq Spin chain models

%---------------------------------------------------------------------------------------------------
%---------------------------------------------------------------------------------------------------

\begin{abstract}
The thermalization phenomenon and many-body quantum statistical properties 
are studied on the example of several observables in isolated spin-chain systems, 
both integrable and generic non-integrable ones. While diagonal matrix elements for 
non-integrable models comply with the eigenstate thermalization hypothesis, 
the integrable systems show evident 
deviations and similarity to properties of noninteracting many-fermion models.
The finite-size scaling reveals that the crossover between two regimes
is given by a scale closely related to the scattering length.  Low-frequency off-diagonal
matrix elements related to d.c.~transport quantities  also
follow in a generic system a behavior analogous to the eigenstate thermalization hypothesis, however
unrelated to the one of diagonal matrix elements.
\end{abstract}

%---------------------------------------------------------------------------------------------------
%---------------------------------------------------------------------------------------------------
\maketitle
\section{Introduction}
Many-body quantum systems and models have been extensively studied in the 
last decades in connection with novel materials, offering a fresh view on the
fundamentals and the interpretation of statistical mechanics. The systematic analysis 
of the phenomena of thermalization and the limitations of a statistical treatment within
isolated \MBQ systems have been recently motivated by experiments on cold atoms in optical
lattices, revealing very slow relaxation to thermal equilibrium \cite{trotz07,cazal00}, but
as well by prototype integrable \MBQ systems as the one-dimensional Heisenberg model realized
in real materials \cite{hess07}.

Specific for lattice \MBQ systems discussed in the above connection  is (in contrast to 
single-body quantum systems)  the exponential growth of the Hilbert space and the number 
of eigenstates with the lattice size $L$. Here, one of the fundamental questions
is to what extent even a single eigenstate or a single chosen initial wave-function could be
the representative of the canonical ensemble average within the given system, both for
static and dynamical quantities. For generic \MBQ systems one of the central statements is
the eigenstate thermalization hypothesis (ETH) \cite{deut91,sred94} 
that for a few-body observable $A$ diagonal matrix elements $A_{nn}$ at a
given energy show only exponentially (in $L$) small deviations from the average, being a
smooth function of the energy only.  Since at the same time the off-diagonal \ME are as well 
exponentially small, the time-average of the observable is determined by diagonal terms
only. Therefore for any initial wave-function with a small energy
uncertainty the long-time average is also equal to the thermal average, this being the
general condition for the quantum thermalization process \cite{rigol12}. We note that such a hypothesis 
is also underlying some numerical methods for the calculation of finite-temperature 
properties, in particular  the microcanonical Lanczos method \cite{long03,prel11}
for $T>0$ static and dynamical properties of lattice \MBQ systems. It seems also
evident that the ETH is intimately related to general properties of eigenenergy spectra, i.e.~level
statistics and dynamics in  generic \MBQ  systems, which reveal Wigner-Dyson level statistics
with the origin in level repulsion and analogy to random matrix spectra \cite{wilk88,acker92}.

The deviations from the ETH and normal thermalization have been detected in several 
directions. The hypothesis is not obeyed in integrable \MBQ systems \cite{rigol07, rigol08,rigol09,
cass11,rigol12}, although some observables can still thermalize, i.e., approach the equilibrium
(canonical ensemble average) value, in particular if the Gibbs statistical ensemble is 
generalized to include all local conserved  quantities in this case \cite{rigol07,cass11}.
The thermalization can become very slow and the validity of ETH can become restricted 
if an initial state is far from equilibrium \cite{rigol08, biro10, genw10, sant11} as
relevant for sudden quenches in cold-atom systems. The latter question is intimately
related to the deviation from integrability \cite{rigol09} and the size of isolated \MBQ systems 
\cite{biro10, genw10, rigol12}. On the other hand, the ETH does not resolve the question of
the relation to off-diagonal \ME (even in generic non-integrable systems) which are, e.g.,  
relevant for transport properties and dissipation in the d.c. limit \cite{zotos96,cast96,
herb12}.

In this paper we study the validity of the ETH and thermalization within a quantum
spin-chain system in one dimension, i.e., the antiferromagnetic and anisotropic $S=1/2$ 
Heisenberg model, including integrable and non-integrable cases.  While we confirm in the generic
\NI case  the ETH for diagonal \ME of several local observables, we find large deviations and 
fluctuations for the \I case. In particular, we show that the spread of diagonal \ME can be qualitatively
and even quantitatively understood from the model of noninteracting fermions. With the
aim to resolve the problem of the breakdown of the ETH in finite systems we perform the
finite-size scaling in \NI systems revealing that the crossover from the \I regime to the 
ETH-consistent behavior is determined by a single scale  $L^*$, coinciding with a transport scattering
length. Another finding is that the off-diagonal \ME at low frequency (small difference of corresponding
eigenenergies) and diagonal \ME are not universally related even in \NI systems, hence the ETH does
not directly address the low-frequency dynamics and the d.c.~transport quantities, and the 
generalization of the ETH is necessary.

The paper is organized as follows: In Sec.~\ref{model} we introduce the model and the considered
observables, i.e., ``kinetic'' energy, spin current, and energy current. In Sec.~\ref{DME} we
analyze the distribution of diagonal \ME for \I and \NI cases. We particularly present a systematic
analysis of the distribution widths as a function of system size and observe in the \NI cases a crossover to
ETH-consistent behavior at a certain length scale, which we connect quantitatively to a transport mean
free path. Section~\ref{ODME} is devoted to the relation between off-diagonal and diagonal \ME as well
as the impact of this relation on low-frequency dynamics and d.c.~transport quantities. In Sec.~\ref{conc}
we finally summarize our results.

%---------------------------------------------------------------------------------------------------
\section{Model and Observables}\label{model}
As the prototype model we study in the following the anisotropic
$S=1/2$ Heisenberg model on a chain with $L$ sites  and periodic boundary
conditions,
\begin{eqnarray}
H &=& J\sum_{i=1}^L (S_i^xS_{i+1}^x+S_i^yS_{i+1}^y
+\Delta S_i^zS_{i+1}^z \nonumber \\ 
&+&\Delta_2S_i^zS_{i+2}^z ) \, , \label{ham}
\end{eqnarray}
where $S_i^\alpha$ ($\alpha = x,y,z$) are spin $S = 1/2$ operators at site $i$ and $\Delta$ 
represents the anisotropy. The nearest-neighbor model is an integrable one and we
introduce the next-nearest-neighbor $zz$-interaction with $\Delta_2 \neq 0$ in order to break
its integrability. It should be reminded that via the Jordan-Wigner transformation \cite{jorwi28}
the Hamiltonian (\ref{ham}) can be mapped on the $t$-$V$-$W$ model of interacting spinless fermions
with the hopping $t=J/2$, the nearest-neighbor interaction $V=J\Delta$, and the next-nearest-neighbor
interaction $W=J\Delta_2$. A consequence of the integrability at
$\Delta_2=0$ is the existence of a  macroscopic number of conserved local quantities
and operators $Q_n, n = 1,\dots,L$  commuting with the Hamiltonian, $[Q_n,H] = 0$.
A nontrivial example is $Q_3 = J^E$  representing the energy current
and leading directly to its non-decaying behavior \cite{zotos97,zotos99} and dissipationless
thermal conductivity \cite{hess07}.  

In order to study \ME properties we choose some simple local 
operators involving only few neighboring sites, however, being still a sum over the whole chain.
Evident candidates are nontrivial quantities involving $n=2$ sites, where we consider the ``kinetic'' energy
\begin{equation}
H^\text{kin} = J\sum_{i=1}^L (S_i^xS_{i+1}^x+S_i^yS_{i+1}^y ) \, ,
\end{equation}
containing the first two terms in Eq.~(\ref{ham}), as well as the spin current
\begin{equation}
J^s = J \sum_{i=1}^L ( S_i^x S_{i+1}^y - S_i^y S_{i+1}^x) \, .
\end{equation}
For a representative of $n=3$ operators we consider the energy current
\begin{eqnarray}
J^E& = & J^2 \sum_{i=1}^L [(S_i^x S_{i+2}^y - S_i^y S_{i+2}^x) S_{i+1}^z \nonumber \\
&-& \Delta (S_i^x S_{i+1}^y - S_i^y S_{i+1}^x)(S_{i-1}^z+S_{i+2}^z)]
\, ,
\end{eqnarray}
not including the $\Delta_2$ term. The choice is motivated by different properties of the
considered operators. While $J^{E}$ is a strictly conserved quantity for the \I case, $J^s$ is not,
but still leads to dissipationless (non-decaying) spin transport. Both are current operators
with \ME distributed around the ensemble average $\langle J^{s,E}_{nm} \rangle=0$. On
the other hand, $H^\text{kin}$ has not such a specific property. In the following we present
results reachable via the exact diagonalization of the model, Eq.~(\ref{ham}), on chains up
to $L=20$. The total spin $S^z_\text{tot}=M$ is fixed to $M=-1$ (in order to avoid
``particle-hole'' symmetry) while we consider both, the representative sector with
wavevector $k=2\pi/L$ and the whole $k$-average as well.

\section{Distribution of Diagonal Matrix Elements}\label{DME}
%---------------------------------------------------------------------------------------------------
\begin{figure}[t]
\includegraphics[width=0.9\columnwidth]{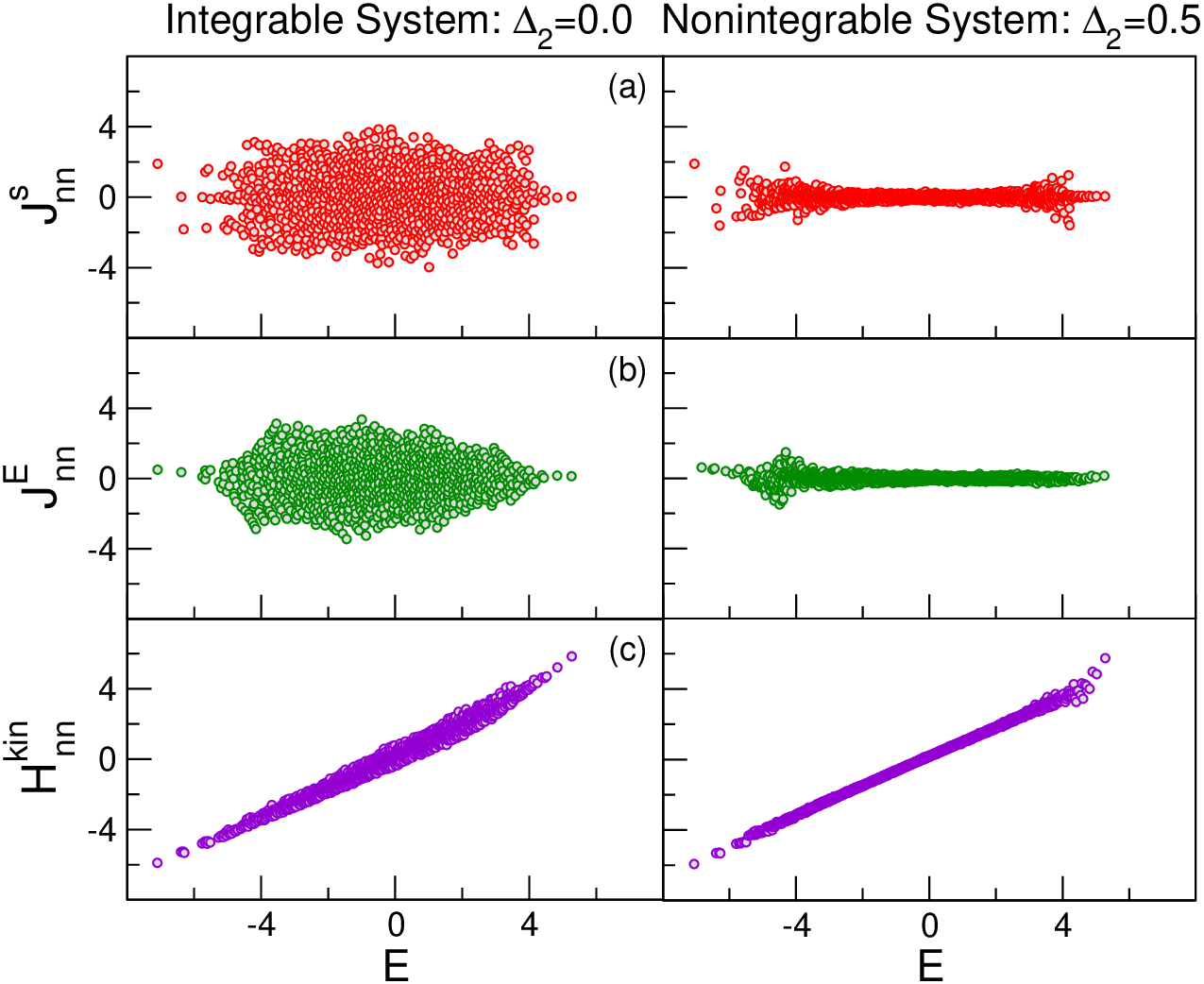}
\caption{(Color online) Distribution of diagonal \ME of (a) spin current
$J^s$, (b) energy current $J^E$, and (c) kinetic energy $H^\text{kin}$
vs.~energy $E$ for the integrable model $\Delta_2=0$ (l.h.s.) and the non-integrable
model $\Delta_2=0.5$ (r.h.s.). In all cases: $\Delta=0.5$, $L=20$, $M=-1$, and
$k=2\pi/L$.} \label{fig1}
\end{figure}
%---------------------------------------------------------------------------------------------------
First, we present results for the distribution of diagonal \ME,
i.e.,~$J^s_{nn}$, $J^E_{nn}$, and  $H^\text{kin}_{nn}$, as they arise varying 
eigenenergies $E=E_n$. In Fig.~\ref{fig1} we show corresponding 2D plots obtained within
the gapless regime ($\Delta=0.5$) and for the magnetization $M=-1$ (due to ``particle-hole'' symmetry 
$J^s_{nn}$ vanishes at $M=0$).  Figure \ref{fig1}
reveals an evident difference between the \NI example with $\Delta_2=0.5$ and the \I case with
$\Delta_2=0$. All quantities  show for the \NI example a narrow distribution around the
average $\langle A_{nn}(E) \rangle$ with the (diagonal) width
\begin{equation}
(\sigma^A_\text{d})^2(E)=\langle A_{nn}(E)^2\rangle-\langle A_{nn}(E)\rangle^2
\end{equation} 
exponentially dependent on the system size $L$ \cite{rigol12}, as later demonstrated in
detail.

On contrary, for the \I case distributions are much wider with a weaker size 
dependence, clearly not obeying the ETH. The distribution for $J^s$ and $J^E$ is
intimately related to the anomalous $T>0$ spin and energy-current stiffness (Drude weight)
for the \I model
\cite{cast96,zotos96,zotos97}, being
within linear response the ballistic contribution to spin and energy conductivity,
\begin{equation}
D^{s,E} (T)=   \frac{\tilde\beta^{s,E}}{L Z} \sum_n {\rm e}^{-\beta E_n} | J^{s,E}_{nn} |^2,
\label{stiff}
\end{equation}
where $\tilde\beta^s=\beta, \tilde\beta^E=\beta^2$ with  $\beta=1/T$.
It is evident that the existence of $D^{s,E}(T>0)>0$ implies that
currents as $J^{s,E}$ do not thermalize to their thermal average $\langle J^{s,E}
\rangle=0$. In particular, their correlation functions do not decay to zero, 
$\langle J^{s,E}(t \to \infty)  J^{s,E} \rangle \neq 0$, and their time evolution depends crucially on the
ensemble of initial states. The same appears to be the case for $H^\text{kin}$, although a physical
interpretation is less familiar. With values of $D^{s,E}(T)$ known from the Bethe
Ansatz \cite{zotos99}, and moreover for the energy-current stiffness $D^E (T\to \infty)$ obtained easily
via the high-$T$ expansion, one can evaluate the distribution widths $\sigma^{s,E}_\text{d}(E)
\propto \sqrt{L}$. 

Since analogous quantities to stiffness are not known in general, one can use in the gapless regime ($\Delta<1$) 
as a semi-quantitative guide results for the $\Delta=0$ model.
The latter can be mapped to the model of non-interacting fermions,
\begin{equation}
H= \sum_k \epsilon_k n_k \, , \quad \epsilon_k = J \cos k \, ,
\end{equation}
being trivially integrable with all $n_k=0,1$ as constants of
motion, with corresponding currents
\begin{equation}
J^s = \sum_k \frac{\partial \epsilon_k}{\partial k} n_k \, , \quad J^E = \sum_k \epsilon_k \frac{\partial \epsilon_k}{\partial k} n_k \, .
\end{equation}
The calculation
of $\sigma^{s,E}_\text{d}(E)$  at fixed magnetization $M=\sum_k (n_k -1/2)$ averaged over 
energies $E$ is for $L \to \infty$ equivalent to the
grand-canonical averaging in the limit $\beta \to 0$ yielding for the unpolarized case $N=L/2$:
$\sigma^{s}_\text{d} = J  \sqrt{L}/\sqrt{8}$ and $\sigma^{E}_\text{d} = J^2 \sqrt{L}/\sqrt{32}$. On the other hand, instead
of $H^\text{kin}$ (being within the $\Delta=0$ limit equal to $H$) one can treat in an analogous way the
complementary potential term $H^\Delta$ with the result $\sigma^\Delta_\text{d} = J\Delta \sqrt{L}/4$ \cite{specificheat}.
We note that the above estimates for the widths $\sigma^{\alpha}_\text{d}$ represent well the numerical
results in Fig.~\ref{fig1} for the \I case with $\Delta > 0$.

%---------------------------------------------------------------------------------------------------
\begin{figure}[t]
\includegraphics[width=0.8\columnwidth]{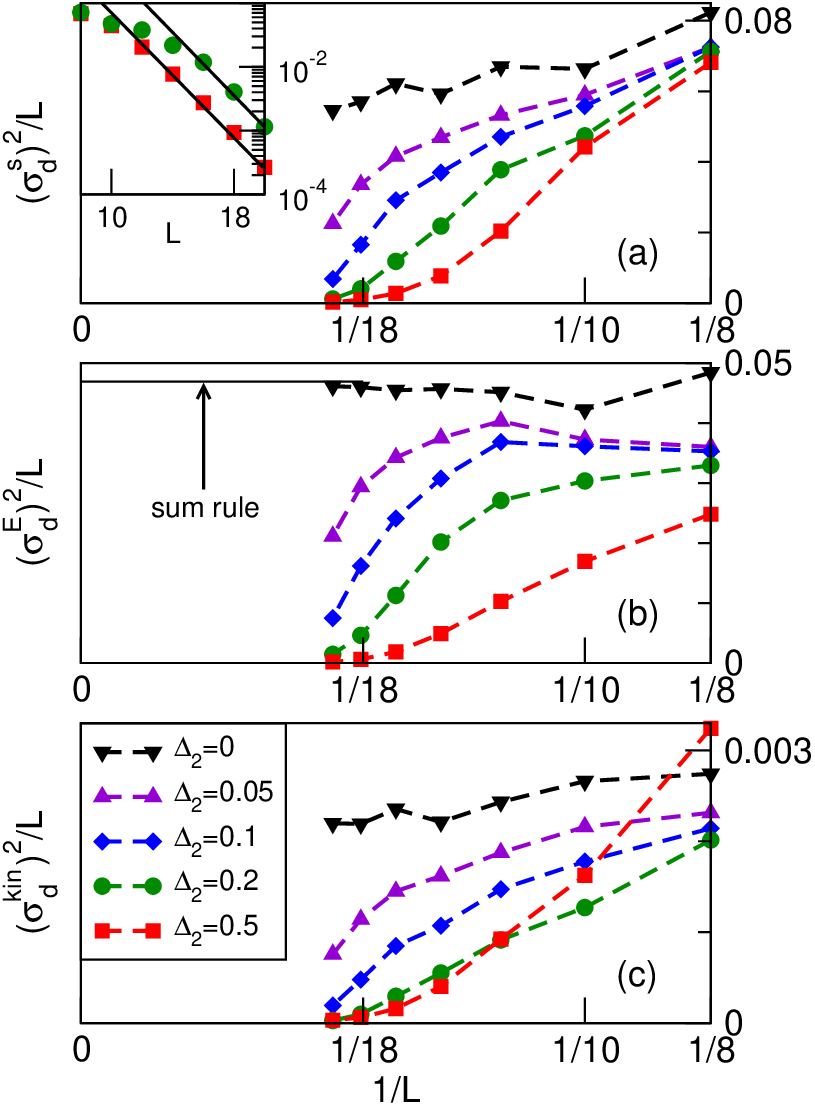}
\caption{(Color online) Finite-size dependence of the diagonal \ME fluctuations of $J^s$, 
$J^E$, and $H^\text{kin}$, respectively, for $\Delta = 0.5, M = -1$ and different 
$\Delta_2$. Fluctuations are evaluated within $E = [-1,1]$. In (b) the exact sum rule is 
indicated (solid line). Inset in (a): Curves for $\Delta_2 = 0.2$ and $0.5$,
illustrating the onset of an exponential decrease with $L$.}
\label{fig2}
\end{figure}
%---------------------------------------------------------------------------------------------------
Next we investigate the crossover from an \I to a \NI system obeying the ETH.
In a finite system fluctuations $\tilde{\sigma}^\alpha_\text{d} = \sigma^\alpha_\text{d}/\sqrt{L}$ with 
$\alpha=(s,E,\text{kin})$ are expected to decrease by introducing 
the \NI perturbation $\Delta_2 \neq 0$. In Fig.~\ref{fig2} we present corresponding
results obtained for different $\Delta_2 =0, \ldots, 0.5$ and sizes $L = 8, \ldots, 20$. In
order to reduce the influence of the energy window, we evaluate the fluctuations
$\sigma^\alpha_\text{d}$ in the range $E=[-1,1]$ and average over all $k$-sectors. 
For the \I case $\Delta_2=0$
the $1/L$-scaling indicates finite values $\tilde \sigma^\alpha_\text{d}(L \to \infty)$. This
coincides with the well defined and nontrivial $D^{s,E}(T \to \infty)$. In particular
$D^E(T \to \infty)/\tilde\beta^{s,E}$ and $\tilde \sigma^E_\text{d}$  can be related to the high-$T$ sum rule  
$(\tilde \sigma^E_\text{d})^2 =(1 + 2 \Delta^2)/32$ \cite{zotos99}. This is, however, not the case for
the \NI case $\Delta_2 \neq 0$. Here, there is an evident decrease with $L$
and crossover to an exponential decrease with $L$, i.e., ETH-consistent behavior above the
crossover scale $L> L^*$.  $L^*$ crucially depends on the perturbation strength $\Delta_2$, but
apparently is roughly the same for all considered quantities.

In the case of currents the ``thermalization length'' $L^*$ may be plausibly interpreted in 
terms of the transport mean free path. The latter can be determined by a standard 
hydrodynamic relation, $1/(q^2 {\cal D}) \gg 1/\gamma$ \cite{robin11}, involving the diffusion 
constant $\cal D$ and the current scattering rate $\gamma$. Identifying the mean free path as
$L^* \approx \pi/q$ then yields
\begin{equation}
L^* \approx \pi \sqrt{\frac{\cal D}{\gamma}}\, .
\end{equation}
For instance, in the case of the spin current, using for $\Delta = 0.5$ and $\Delta_2 = 0.2 [0.5]$
the known quantitative values \cite{herb12}
${\cal D}^s=2.1 [3.6]$ and  $\gamma^s=0.23 [0.13]$ at
$\beta \rightarrow 0$, one finds $L^* \approx 10 [16]$. This value turns out to agree well with the scale observed
in the inset of Fig.~\ref{fig2}. Moreover, $\gamma^s \rightarrow 0$ as $\Delta_2 \rightarrow 0$
is consistent with a diverging $L^*$.

%---------------------------------------------------------------------------------------------------
\section{Relation between Off-diagonal and Diagonal Matrix Elements}\label{ODME}
\begin{figure}[t]
\includegraphics[width=0.76\columnwidth]{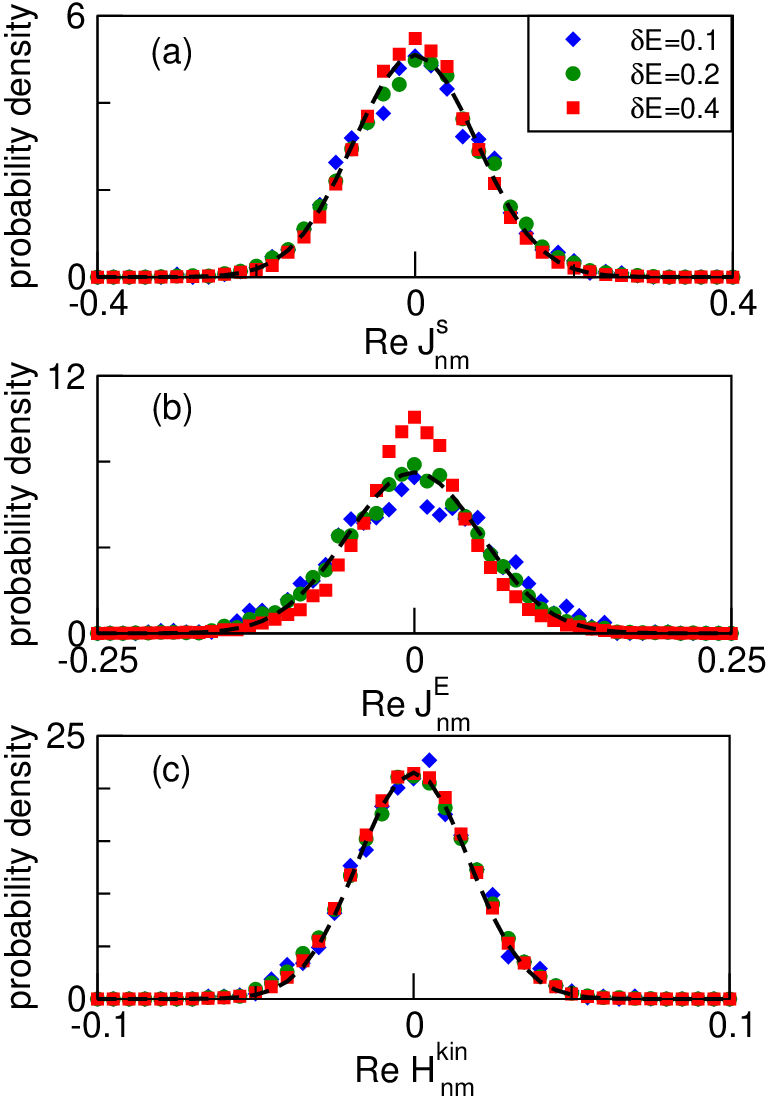}
\caption{(Color online)  Probability distribution for the real part of the off-diagonal \ME
of $J^s$, $J^E$, and $H^\text{kin}$, respectively, for a \NI model with 
$\Delta = \Delta_2 = 0.5$, $L=18$, and $M=-1$. For comparison, 
Gaussian functions are indicated (dashed curves).} \label{fig3}
\end{figure}

\begin{figure}[t]
\includegraphics[width=0.8\columnwidth]{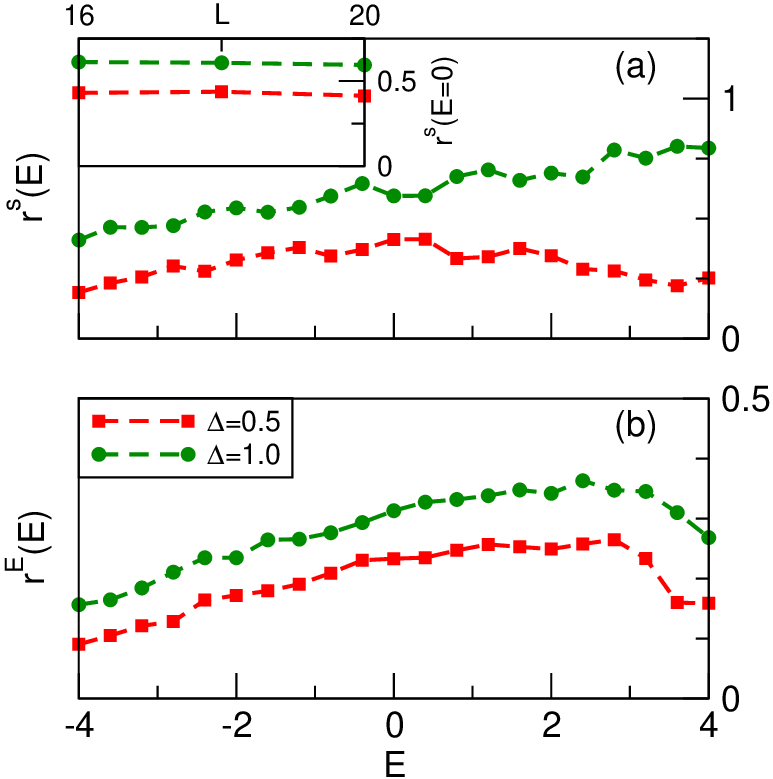}
\caption{(Color online) Ratio between off-diagonal and diagonal \ME fluctuations
for $J^s$ and $J^E$, respectively, for a \NI model with $\Delta =0.5,1 $, $\Delta_2 = 0.5$, $L=20$,
$M=-1$, and $k$-average. Inset in (a): Finite-size dependence of $r^s(E=0)$.}
\label{fig4}
\end{figure}

Finally, let us address the relation between off-diagonal and diagonal \ME. Since for the 
\I system the behavior can be very singular \cite{herb12}, we concentrate on the
generic \NI cases satisfying the ETH. In Fig.~\ref{fig3} we present the probability distribution
of off-diagonal \ME, e.g., $\text{Re} J^{s,E}_{nm}$ and $\text{Re} H^\text{kin}_{nm}$,
evaluated for $\Delta=0.5$, $\Delta_2=0.5$ in the energy window $E_n, E_m = [-\delta E/2,
\delta E/2]$ with various $\delta E \ll J$. Using a small $\delta E$ respects the topology of
a banded random matrix \cite{cohen01} with a band width on the order of the exchange coupling constant
$J$. Resulting distributions do clearly not depend on $\delta E$ and appear to be Gaussian with
zero mean.
%---------------------------------------------------------------------------------------------------
%---------------------------------------------------------------------------------------------------
It is a nontrivial question whether the fluctuations of off-diagonal and diagonal \ME follow the same
scaling with $L$. It is therefore important to investigate the ratio of off-diagonal and
diagonal \ME fluctuations
\begin{equation}
r^{\alpha}(E)= \frac{(\sigma^{\alpha}_\text{od})^2(E)}{(\sigma^{\alpha}_\text{d})^2(E)}\, , \quad
(\sigma_\text{od})^2(E) = \langle |A_{mn}(E)|^2 \rangle \, .
\end{equation}
Results for the spin and energy current are presented
in Fig.~\ref{fig4}, shown vs.~$E$ for $\Delta_2=0.5$ and $\Delta=0.5, 1.0$.
They  indicate that $r^\alpha(E)$  is not universal (depends on $\alpha$ and 
model parameters) and smoothly varies with $E$, but most important is 
the independence of $L$. We can conclude that for the cases considered  here $r^\alpha$ are
not following relations within the random-matrix theory \cite{wilk88,cast96} implying
generally $r = 1/2$ for the Gaussian Orthogonal Ensemble (and $r = 1$ for the
Gaussian Unitary Ensemble). On the other hand, the ratio still remains within an order of 
magnitude in contrast to the \I case where in the gapless regime the ratio appears to vanish
leaving finite only diagonal \ME \cite{herb12}.

The above observation becomes relevant in the evaluation of d.c.~transport quantities,
which are within linear response theory related to the low-$\omega$ absorption 
\cite{maha}, e.g.,~the spin conductivity (diffusivity) and thermal conductivity, respectively, are
in analogy to Eq.~(\ref{stiff}),
\begin{equation}
C^\alpha(\omega) = \frac{\tilde \beta^\alpha \pi}{LZ} \sum_{m \neq n} {\rm e}^{-\beta E_n} | 
J^\alpha_{mn} |^2  \delta(\omega- E_m+E_n) \, , \label{sig}
\end{equation}
where the d.c.~limit should be considered as $C^\alpha_0= C^\alpha(\omega \to
0)$ and can be expressed as 
\begin{equation}
C^\alpha_0 = \frac{\tilde \beta^\alpha \pi}{Z}  \int e^{-\beta E} \rho^2(E) (\tilde
\sigma^\alpha_\text{od})^2(E) dE \, ,
\label{sig0}
\end{equation}
where $\rho(E)$ is the \MBQ density of states. From our analysis it follows that
in general $\tilde \sigma^\alpha_\text{od}(E)$ cannot be represented by diagonal
$\tilde \sigma^\alpha_\text{d}(E)$, although the qualitative behavior appears closely related (and
even quantitative for $C^s_0$ as evident from Fig.~\ref{fig4}a).
Note that for the case of  $J^s$ diagonal \ME can be also expressed as the sensitivity of many-body 
levels to a fictitious flux $\phi$ (or boundary conditions),
i.e.~$J^s_{nn} \propto \partial E_n /\partial \phi$, and the latter relation has been previously
employed to evaluate the d.c.~transport in, e.g., disordered systems \cite{edwa72,acker92}.

%---------------------------------------------------------------------------------------------------

%---------------------------------------------------------------------------------------------------
\section{Conclusions}\label{conc}
Let us in conclusion summarize our results, which may be generic beyond
spin-chain systems. The behavior of the considered \NI systems we find consistent 
with the ETH for all considered quantities.
If we consider the time evolution of an observable, it can be in terms of (finite-system)
eigenstates represented as
\begin{eqnarray}
A(t) &=& \langle \Psi(t)| A| \Psi(t)  \rangle = \sum_n |c_n|^2   A_{nn} + \nonumber \\
&+&   \sum_{n \neq m}  c^*_n c_m {\rm e}^{\imath(E_n-E_m)t} A_{nm}. \label{at}
\end{eqnarray}
In a system obeying ETH, the off-diagonal contribution vanishes for long times $t \to \infty$,
due to the exponential smallness of off-diagonal \ME (compare insets of Fig.~\ref{fig2} and Fig.~\ref{fig4})
as well due to dephasing \cite{rigol12}. 
If the initial state $|\Psi_0\rangle $ is a microcanonical one with a narrow distribution
$\delta E$ [with $(\delta E)^2 = \sum_n |c_n|^2 (E_n - \bar E)^2$], and due to ETH 
$ A_{nn}  \sim \langle A \rangle (\bar E)$,   the first term leads to the 
microcanonical  average  $A(t) = \langle A \rangle (\bar E) $  in a large system
coinciding with the canonical thermodynamical average at a finite $T>0$, 
where $E(T) = \bar E$. Such a scenario is then consistent with the ``normal'' 
quantum  thermalization.

In an \I spin chain the distribution of diagonal \ME
is large, the long-time average [still neglecting off-diagonal terms in Eq.~(\ref{at})]
in general depends on $|\Psi_0\rangle $ and corresponding  $c_n$, even for a small
energy uncertainty $\delta E$. In order to satisfy $A(t \to \infty)  = \langle A \rangle $
one needs assumptions on the distribution of coefficients $c_n$. E.g., in a large enough system
randomly chosen $c_n$ would plausibly be adequate. In fact, the microcanonical Lanczos method 
for the evaluation of $T>0$ properties \cite{long03,prel11}, based on the microcanonical states 
and the Lanczos procedure, contains such a choice achieved by random sampling. 
Hence, a random microcanonical state in a large \MBQ system would mostly obey 
the thermalization process. Still, this is not at all the case for particular states as, e.g., 
reached by (strong) quenching in an \I system, but as well not in a generic system \cite{rigol09,biro10} 
since the initial state after the quench is not necessarily the 
microcanonical one with small $\delta E$. 

Analyzing the extent of the validity of the ETH and thermalization in a finite-size \MBQ system, 
we find effectively that perturbed \I systems 
beyond the crossover length $L^*$ behave as generic \NI ones. Since in a ``normal'' spin
system only total spin and energy are conserved, one can design
two relevant diffusion scales and plausibly the largest would determine
$L^*$, which then appears to dominate the scaling of all quantities, as shown in Fig.~\ref{fig3}.
The understanding and the determination of $L^*$ is evidently an important theoretical goal,
relevant also for experiments dealing with systems close to integrability \cite{hess07,
hlub12}.

The ETH addresses thermalization and statistical description of static quantities 
in \MBQ systems, with the behavior determined by diagonal \ME.
On the other hand, d.c.~transport quantities and low-$\omega$ dynamics
involve only off-diagonal \ME. We note that in a generic system, 
properties analogous to the ETH can be defined for off-diagonal \ME close in energy, 
in particular obeying the Gaussian distribution and exponential dependence on size. Also, the relation
between diagonal and off-diagonal \ME is independent of size $L$, but still the ratio is
not universal. In this sense, our results show that for such considerations the
generalization of the ETH is needed but also is straightforward, and it can
include the response to weak external fields and dissipation phenomena in \MBQ systems.

%---------------------------------------------------------------------------------------------------

\acknowledgements This research was supported by the RTN-LOTHERM project and the Slovenian Agency 
grant No. P1-0044.
%---------------------------------------------------------------------------------------------------

\end{document}